\documentstyle[12pt]{article}

\let\ni=\noindent

\begin{document}

\baselineskip 0.75cm
 
\pagestyle {plain}

\setcounter{page}{1}

\renewcommand{\thefootnote}{\fnsymbol{footnote}}

\newcommand{\CKM}{Cabibbo---Kobayashi---Maskawa }

\newcommand{\SM }{Standard Model }

\newcommand{\UK}{Super--Kamiokande }

\newcommand{\onu }{\stackrel{\circ}{\nu} }

\newcommand{\mnu }{\nu_s^{(\mu)} }

\newcommand{\enu }{\nu_s^{(e)} }

\pagestyle {plain}

\setcounter{page}{1}

\pagestyle{empty}

~~~
\hfill IFT/99-24

\vspace{0.3cm}

\renewcommand{\thefootnote}{\fnsymbol{footnote}}

{\large\centerline{\bf Option of three pseudo--Dirac neutrinos{\footnote{
Supported in part by the Polish KBN--Grant 2 P03B 052 16 (1999--2000).}}}}

\vspace{0.8cm}

{\centerline {\sc Wojciech Kr\'{o}likowski}}

\vspace{0.8cm}

{\centerline {\it Institute of Theoretical Physics, Warsaw University}}

{\centerline {\it Ho\.{z}a 69,~~PL--00--681 Warszawa, ~Poland}}

\vspace{0.5cm}

{\centerline{\bf Abstract}}

 As an alternative for popular see--saw mechanism, the option of three pseudo%
--Dirac neutrinos is discussed, where $\frac{1}{2}(m^{(L)} + m^{(R)}) \ll m^{(
D)}$ for their Majorana and Dirac masses. The actual neutrino mass matrix is 
assumed in the form of tensor product $ M^{(\nu)} \otimes {\left( \begin{array}
{cc} \lambda^{(L)} & 1 \\ 1 & \lambda^{(R)} \end{array} \right)}$, where $ M^{(
\nu)}$ is a neutrino family mass matrix ($ M^{(\nu)\,\dagger} = M^{(\nu)}$) 
and $\lambda^{(L,R)} \equiv m^{(L,R)}/m^{(D)}$ with $ m^{(L)}$, $ m^{(R)}$ and 
$ m^{(D)} $ being taken as universal for three neutrino families. It is shown 
that three neutrino effects (deficits of solar $\nu_e $'s and atmospheric $
\nu_\mu $'s  as well as the possible LSND excess of $\nu_e $'s in accelerator 
$\nu_\mu $ beam) can be nicely described by the corresponding neutrino 
oscillations, though the LSND effect may, alternatively, be eliminated (by a 
parameter choice). Atmospheric $\nu_\mu $'s oscillate dominantly into $\nu_\tau
$'s, while solar $\nu_e $'s --- into (existing here automatically) Majorana 
sterile counterparts of $\nu_e $'s. A phenomenological texture for neutrinos, 
compatible with the proposed description, is briefly presented.

\vspace{0.2cm}

\ni PACS numbers: 12.15.Ff , 14.60.Pq , 12.15.Hh .

\vspace{0.5cm}

\ni October 1999

\vfill\eject

~~~~
\pagestyle {plain}

\setcounter{page}{1}

\setcounter{footnote}{6}


 As is well known, in the popular see--saw mechanism [1] righthanded neutrinos 
get (by assumption) large Majorana--type masses and become practically 
decoupled from lefthanded neutrinos that are allowed to carry only small 
Majorana--type masses. On the other hand, in such a case, oscillations of three
neutrinos can hardly explain all three neutrino effects (deficits of solar $
\nu_e $'s and atmospheric $\nu_\mu $'s  as well as the possible LSND excess of 
$\nu_e $'s in accelerator $\nu_\mu $ beam). This may suggest the existence of, 
at least, one extra neutrino called sterile ({\it i.e.}, passive to all \SM 
gauge interactions), mixing with three active neutrinos [2].

The option of three pseudo--Dirac neutrinos [3], where the Dirac mass $m^{(D)}$
dominates over two Majorana masses $m^{(L)}$ and $m^{(R)}$ (within their 
Majorana--type masses), is orthogonal to the see--saw mechanism with $m^{(R)}$
dominating over $m^{(L)}$ and $m^{(D)}$, and $m^{(D)}$ over $m^{(L)}$. Thus, in
contrast to the see--saw, this option cannot guarantee automatically small 
Majorana--type masses for lefthanded neutrinos (their smallness must be here 
directly assumed). However, as will be shown in this note, oscillations of 
three pseudo--Dirac neutrinos are sufficient to explain all three neutrino 
effects without introducing any extra sterile neutrinos. This is due to the 
automatic existence of three conventional Majorana sterile neutrinos $\nu^{(s)
}_\alpha \equiv \nu_{\alpha R} + (\nu_{\alpha R})^c $ which, in the pseudo--%
Dirac case, are not decoupled from three conventional Majorana active neutrinos
$\nu^{(a)}_\alpha \equiv \nu_{\alpha L} + (\nu_{\alpha L})^c$ ($\alpha = e\,,\,
\mu\,,\,\tau $).

 Thus, let us consider three flavor neutrinos $\nu_e $, $\nu_\mu $, $\nu_\tau $
and , approximately, assume for them the mass matrix in the form of tensor product
of the neutrino family $ 3\times 3 $ mass matrix $ \left(M^{(\nu)}_{\alpha 
\beta}\right)\;\;(\alpha,\beta = e,\,\mu,\,\tau)$ and the Majorana $2\times 2$ 
mass matrix

\vspace{-0.1cm}

\begin{equation}
\left( \begin{array}{cc} m^{(L)} & m^{(D)} \\ m^{(D)} & m^{(R)} \end{array} 
\right)\;\;,
\end{equation}

\vspace{-0.1cm}

\ni the latter divided by $ m^{(D)}$ (with $ m^{(D)}$ included into $ M^{(\nu)
}_{\alpha \beta}$). Then, the neutrino mass term in the lagrangian gets the 
form

\vspace{-0.2cm}

\begin{eqnarray}
-{\cal L}_{\rm mass} & = & \frac{1}{2}\sum_{\alpha \beta} \left(\overline{\onu
}^{(a)}_\alpha\;,\;\overline{\onu}^{(s)}_\alpha \right)\;{M}^{(\nu)}_{\alpha 
\beta} \; \left( \begin{array}{cc} \lambda^{(L)} & 1 \\ 1 & \lambda^{(R)} 
\end{array} \right)\;\left(\begin{array}{c} \onu^{(a)}_\beta \\ \onu_\beta^{(s)
} \end{array} \right) \nonumber \\ & = & \frac{1}{2}\sum_{\alpha \beta} \left(
\overline{\left( \onu_{\alpha L} \right)^c} \;,\; \overline{\onu_{\alpha R}} 
\right) \; {M}^{(\nu)}_{\alpha \beta}\;\left(\begin{array}{cc}\lambda^{(L)} & 1
\\ 1 & \lambda^{(R)} \end{array} \right)\;\left( \begin{array}{c} \onu_{\beta L
} \\ \left( \onu_{\beta R}\right)^c \end{array} \right) + {\rm h.c.}\;,
\end{eqnarray}


\ni where


\begin{equation}
\onu^{(a)}_\alpha \equiv\; \onu_{\alpha\,L} + \left(\onu_{\alpha\,L}\right)^c
\,,\,\onu^{(s)}_\alpha \equiv\; \onu_{\alpha\,R} + \left(\onu_{\alpha\,R}
\right)^c
\end{equation}

\ni and $\lambda^{(L,R)} \equiv m^{(L,R)}/m^{(D)}$. Here, $\onu^{(a)}_\alpha $ 
and $\onu_\alpha^{(s)}$ are the conventional Majorana active and sterile neutr%
inos of three families as they appear in the lagrangian before diagonalization 
of neutrino and charged--lepton family mass matrices. Due to the relation $
\overline{\nu^c_\alpha}\nu_\beta = \overline{\nu^c_\beta} \nu_\alpha $, the 
neutrino family mass matrix $ M^{(\nu)} = M^{(\nu)\,\dagger}$, when standing at
the position of $\lambda^{(L)}$ and $\lambda^{(R)}$ in Eq. (2), reduces to its 
symmetric part $\frac{1}{2}(M^{(\nu)} + M^{(\nu)\,T} ) $ equal to its real part
$\frac{1}{2}(M^{(\nu)} + M^{(\nu)\,*}) = {\rm Re}\, M^{(\nu)} $. We will simply
assume that (at least approximately) $ M^{(\nu)} = M^{(\nu)\,T} = M^{(\nu)\,*}
$, and hence for neutrino family diagonalizing matrix $ U^{(\nu)} = U^{(\nu)\,
*} = \left(U^{(\nu)\,-1}\right)^T $. Then, CP violation for neutrinos does not 
appear if, in addition, for charged--lepton diagonalizing matrix $ U^{(e)} = 
U^{(e)\,*}$. Further on, we will always assume that $ 0 < \frac{1}{2}(\lambda^{(L)} 
+ \lambda^{(R)})\;\;(\equiv \lambda^{(M)}$) $\ll 1 $ (the pseudo--Dirac option,
in contrast to the see--saw mechanism, where $\lambda^{(L)} \ll 1 \ll 
\lambda^{(R)}) $.

 Then, diagonalizing the neutrino mass matrix, we obtain from Eq. (2) 


\begin{equation}
-{\cal L}_{\rm mass} = \frac{1}{2} \sum_i \left( \overline{\nu}^I_i\;,
\; \overline{\nu}_i^{II} \right)\;m_{\nu_i}\;\left( \begin{array}{cc} 
\lambda^I & 0 \\ 0 & \lambda^{II} \end{array} \right)\;\left(\begin{array}{c} 
\nu^{I}_i \\ \nu^{II}_i \end{array} \right)\;\;,
\end{equation}


\ni where


\begin{equation}
\left( {U}^{(\nu)\,\dagger}\right)_{i\,\alpha} {M}^{(\nu)}_{\alpha \beta} {U}^{
(\nu)}_{\beta\,j} = m_{\nu_i} \delta_{ij}\;\;\;,\;\;\;\lambda^{I,\,II} \simeq 
\mp 1 + \lambda^{(M)} \simeq \mp 1
\end{equation}


\ni $(i,\,j = 1,2,3)$ and


\begin{equation}
\nu_i^{I,\,II} \simeq \sum_i \left({U}^{(\nu)\,\dagger}\right)_{i\,\alpha} 
\frac{1}{\sqrt{2}}\left(\onu^{(a)}_\alpha \mp \onu^{(s)}_\alpha\right) =
\sum_i {V}_{i\,\alpha} \frac{1}{\sqrt{2}}\left(\nu^{(a)}_\alpha \mp \nu^{(s)
}_\alpha\right)
\end{equation}


\ni with $ V_{i\,\alpha} = \left({U}^{(\nu)\,\dagger}\right)_{i\,\beta} U^{(e)
}_{\beta \alpha}$ describing the lepton counterpart of the \CKM matrix. Here,


\begin{equation}
\nu_\alpha^{(a,s)} \equiv \nu_{\alpha\,L,R} + \left(\nu_{\alpha\,L,R}\right)^c 
= \sum_\beta \left( U^{(e)\,\dagger}\right)_{\alpha \beta} \onu^{(a,s)}_\beta 
\simeq \sum_i \left( V^\dagger \right)_{\alpha\,i} \frac{1}{\sqrt{2}}\left(\pm 
\nu^I_i + \nu^{II}_i \right) 
\end{equation}


\ni and


\begin{equation}
\left( U^{(e)\,\dagger}\right)_{\alpha \gamma} M^{(e)}_{\gamma \delta} U^{(e)
}_{\delta \beta} = m_{e_\alpha} \delta_{\alpha \beta}\;,
\end{equation}

\ni where $ \left(M^{(e)}_{\alpha \beta}\right)\;\;(\alpha,\beta = e,\,\mu,\,
\tau)$ is the mass matrix for three charged leptons $e^-,\,\mu^-,\,\tau^- $, 
giving their masses $m_e,\,m_\mu,\,m_\tau $ after its diagonalization is 
carried out. Now, $\nu_\alpha^{(a)}$ and $\nu_\alpha^{(s)}$ are the convent%
ional Majorana active and sterile flavor neutrinos of three families, while $
\nu_i^I $ and $\nu_i^{II}$ are Majorana massive neutrinos carrying masses 
$m_{\nu_i}\lambda^{I}$ and $m_{\nu_i}\lambda^{II}$ (phenomenologically, they
get nearly degenerate masses $|m_{\nu_i}\lambda^{I}|$ and $|m_{\nu_i}\lambda^{
II}|$).

 If CP violation for neutrinos does not appear or can be neglected, the 
probabilities for oscillations $\nu_\alpha^{(a)} \rightarrow \nu_\beta^{(a)}$ 
and $\nu_\alpha^{(a)} \rightarrow \nu_\beta^{(s)}$ are given by the following 
formulae (in the pseudo--Dirac case):


\begin{eqnarray}
\lefteqn{P\left(\nu^{(a)}_\alpha \rightarrow \nu^{(a)}_\beta\right)\! =
\!|\langle \nu^{(a)}_\beta |e^{i P L}|\nu^{(a)}_\alpha \rangle |^2 = \delta_{
\beta\,\alpha} - \sum_i |V_{i\,\beta}|^2 |V_{i\,\alpha}|^2 \sin^2 \left( x^{II
}_i \! -\! x^I_i \right)} \nonumber \\ & & \!\!\! -\! \sum_{j>i} V_{j\,\beta} 
V^*_{j\,\alpha} V^*_{i\,\beta} V_{i\,\alpha}\!\left[ \sin^2 \left(x^I_j \! -\! 
x^I_i \right) + \sin^2 \left(x^{II}_j \! -\! x^{II}_i \right) + \sin^2 \left(
x^{II}_j\! - \! x^I_i \right) + \sin^2 \left( x^I_j \! -\! x^{II}_i \right) 
\right] \nonumber \\ & &
\end{eqnarray}


\ni and


\begin{eqnarray}
\lefteqn{P\left(\nu^{(a)}_\alpha \rightarrow \nu^{(s)}_\beta\right)\! =\!
|\langle \nu^{(s)}_\beta |e^{i P L}|\nu^{(a)}_\alpha \rangle |^2\! =\! \sum_i
|V_{i\,\beta}|^2 |V_{i\,\alpha}|^2 \sin^2 \left( x^{II}_i \! -\! x^I_i \right)}
\nonumber \\ & & \!\!\! -\!\sum_{j>i}\! V_{j\,\beta}V^*_{j\,\alpha}V^*_{i\,
\beta}V_{i\,\alpha}\left[ \sin^2 \left( x^I_j \! -\! x^I_i \right)\! + \!
\sin^2 \left(x^{II}_j \! -\! x^{II}_i \right)\! - \!\sin^2 \left( x^{II}_j\! 
-\! x^I_i \right)\! - \!\sin^2 \left(x^I_j\! -\! x^{II}_i \right) \right], 
\nonumber \\ & &
\end{eqnarray}


\ni where $ P |\nu^{I,\,II}_i \rangle = p^{I,\,II}_i |\nu^{I,\,II}_i \rangle 
$ , $p_i^{I,\,II} = \sqrt{E^2 - (m_{\nu_i}\lambda^{I,\,II})^2} \simeq E - 
(m_{\nu_i}\lambda^{I,\,II})^2/2E $ and


\begin{equation}
x_i^{I,\,II} = 1.27\frac{ (m_{\nu_i}^2\lambda^{I,\,II})^2 L}{E}\;\;,\;\;
(\lambda^{I,\,II})^2 = 1 \mp 2\lambda^{(M)} \simeq 1
\end{equation}


\ni with $ m_{\nu_i}$, $ L $ and $ E $ expressed in eV, km and GeV, respect%
ively ($ L $ is the experimental baseline). Here, due to Eqs. (11),


\begin{equation}
x_i^{II} - x_i^I = 1.27\frac{4 m_{\nu_i}^2\lambda^{(M)} L}{E}
\end{equation}


\ni and for $ j > i $


\begin{equation}
x^I_j  - x^I_i  \simeq x^{II}_j  - x^{II}_i \simeq  x^{II}_j - x^I_i \simeq 
x^I_j - x^{II}_i \simeq 1.27\frac{( m_{\nu_j}^2 -  m_{\nu_i}^2) L}{E}\;.
\end{equation}


\ni Then, the bracket [ ] in Eq. (9) and (10) is reduced to $ 4 \sin^2 1.27
(m_{\nu_j}^2 - m_{\nu_i}^2) L/E $ and 0, respectively. The probability sum rule
$\sum_\beta \left[P\left(\nu^{(a)}_\alpha \rightarrow \nu^{(a)}_\beta\right) + 
P\left(\nu^{(a)}_\alpha \rightarrow \nu^{(s)}_\beta \right) \right] = 1 $ 
follows readily from Eqs. (9) and (10).

 Notice that in the case of lepton \CKM matrix being nearly unit, $\left( V_{i
\,\alpha}\right) \simeq \left(\delta_{i\,\alpha}\right)$, the oscillations $
\nu_\alpha^{(a)} \rightarrow \nu_\beta^{(a)}$ and $\nu_\alpha^{(a)} \rightarrow
\nu_\beta^{(s)}$ are essentially described by the formulae


\begin{eqnarray}
P\left(\nu^{(a)}_\alpha \rightarrow \nu^{(a)}_\beta\right) & \simeq &
\delta_{\beta \alpha} - P\left(\nu^{(a)}_\alpha \rightarrow 
\nu^{(s)}_\beta \right) \;, \nonumber \\ P\left(\nu^{(a)}_\alpha \rightarrow 
\nu^{(s)}_\beta\right) & \simeq & \delta_{\beta \alpha} \sin^2 \left(
1.27\frac{4 m_{\nu_\alpha}^2\lambda^{(M)} L}{E} \right)  
\end{eqnarray}


\ni corresponding to three maximal mixings of $\nu^{(a)}_\alpha $ with $\nu^{(s
)}_\alpha $ $(\alpha = e,\,\mu,\,\tau)$. Of course, for a further discussion of
the oscillation formulae (9) and (10), in particular those for appearance modes
$\nu^{(a)}_\alpha \rightarrow \nu^{(a)}_\beta\;\;(\alpha \neq \beta)$, a 
detailed knowledge of $\left( V_{i\,\alpha}\right)$ is necessary. 

 Further on, we will concentrate on the attractive mass hierarchy

\begin{equation}
m_{\nu_1}^2 \ll m_{\nu_2}^2 \simeq m_{\nu_3}^2 
\end{equation}

\ni that may enable us to interpret the LSND scale $\Delta m^2_{\rm LSND}$ as 
$ m^2_{\nu_2} - m^2_{\nu_1}$, while both smaller scales, the solar scale 
$\Delta m^2_{\rm sol}$ and atmospheric scale $\Delta m^2_{\rm atm}$ may be 
equal to $( m_{\nu_1}\lambda^{II})^2 - (m_{\nu_1}\lambda^I)^2 \simeq 4 m_{\nu_1
}^2 \lambda^{(M)}$ and $ m^2_{\nu_3} - m^2_{\nu_2}$, respectively. 

 In fact, due to Eqs. (12), (13) and (15), and the unitarity of $\left( V_{i\,
\alpha}\right)$, the oscillation formulae (9) imply

\vspace{-0.2cm}

\begin{eqnarray} 
P\left(\nu_\alpha \rightarrow \nu_\alpha\right) \simeq & 1 & \!\!-\; |V_{1 
\alpha}|^4 \sin^2 \left(1.27 \frac{4 m_1^2\lambda^{(M)} L}{E} \right) \nonumber
\\ & & \!\! -\; \left( |V_{2 \alpha}|^4 + |V_{3 \alpha}|^4 \right)\sin^2 \left(
1.27 \frac{4 m_2^2 \lambda^{(M)} L}{E} \right) \nonumber \\ & & \!\!-\; 4 |V_{
1 \alpha}|^2 \left( 1 - |V_{1 \alpha}|^2 \right) \sin^2 \left(1.27 \frac{\Delta
m_{21}^2 L}{E} \right) \nonumber \\ & & \!\!-\; 4 |V_{2 \alpha}|^2 |V_{3 \alpha
}|^2 \sin^2 \left( 1.27 \frac{\Delta m_{32}^2 L}{E} \right)
\end{eqnarray}

\ni for $\alpha = e, \mu $ and

\begin{eqnarray}
P\left(\nu_\mu \rightarrow \nu_\beta \right) & \simeq & \!\!-\;|V_{1 \beta}|^2
|V_{1 \mu}|^2 \sin^2 \left( 1.27 \frac{4 m_1^2\lambda^{(M)} L}{E} \right) 
\nonumber \\  & & \!\!-\; \left( |V_{2 \beta}|^2 |V_{2 \mu}|^2 + |V_{3 \beta}
|^2 |V_{3 \mu}|^2 \right) \sin^2 \left( 1.27 \frac{4 m_2^2 \lambda^{(M)} L}{E} 
\right) \nonumber \\ & & \!\!+\; 4 |V_{1 \beta}|^2 |V_{1 \mu}|^2 \sin^2 \left( 
1.27 \frac{\Delta m_{21}^2 L}{E} \right) \nonumber \\ & & \!\!+\; 4 \left( |V_{
2 \beta}|^2 |V_{2 \mu}|^2 + V_{1 \beta} V_{1 \mu}^* V_{2 \beta}^* V_{2 
\mu} \right) \sin^2 \left( 1.27 \frac{\Delta m_{32}^2 L}{E} \right)
\end{eqnarray}

\ni for $\beta = e, \tau $. Here, $ m_i \equiv m_{\nu_i}$ and $\Delta m^2_{j i}
\equiv m^2_j - m^2_i $ (Eq. (15) shows that $\Delta m^2_{21} \simeq m^2_2 $
and $\Delta m^2_{31} \simeq m^2_3 $). Note that $\nu^{(a)}_{\alpha L} \equiv
\nu_{\alpha L}$ and $\nu^{(s)}_{\alpha L} \equiv (\nu_{\alpha R})^c $.

 We intend to relate Eqs. (16) with $\alpha = e $ and $\alpha = \mu $ to the
experimental results concerning the deficits of solar $\nu_e$'s [4] and atmo\-%
spheric $\nu_\mu $'s [5], respectively, and Eq. (17) with $\beta = e $ to the 
possible LSND excess of $\nu_e $'s in accelerator $\nu_\mu $ beam [6].

 To this end let us make the numerical conjecture that

\begin{equation}
1.27 \frac{4 m_{1}^2\lambda^{(M)} L_{\rm sol}}{E_{\rm sol}} = O(1)\;\;,\;\;1.27
\frac{\Delta m_{32}^2 L_{\rm atm}}{E_{\rm atm}} = O(1)\;\;,\;\;
1.27 \frac{\Delta m_{21}^2 L_{\rm LSND}}{E_{\rm LSND}} = O(1)\;,
\end{equation}

\ni and

\begin{equation}
\left( 4 m_{2}^2\lambda^{(M)}\right)^2 \ll \left(\Delta m_{32}^2 \right)^2
\end{equation}

\ni (while not necessarily $4 m_{2}^2\lambda^{(M)} \ll \Delta m_{32}^2 $). 
Then, we get from Eqs. (16) and (17) the following oscillation formulae:

\begin{eqnarray} 
P\left(\nu_e \rightarrow \nu_e\right) \simeq & 1 & \!\!-\; |V_{1 e}|^4 \sin^2 
\left(1.27 \frac{4 m_1^2\lambda^{(M)} L_{\rm sol}}{E_{\rm sol}} \right)
\nonumber \\ & & \!\! -\;\frac{1}{2}\left[|V_{2 e}|^4 + |V_{3 e}|^4 + 4 |V_{
1 e}|^2 \left( 1 - |V_{1 e}|^2 \right) + 4 |V_{2 e}|^2 |V_{3 e}|^2\right]\;, 
\\ P\left(\nu_\mu \rightarrow \nu_\mu\right) \simeq & 1 & \!\!-\; 
4 |V_{2 \mu}|^2 |V_{3 \mu}|^2 \sin^2 \left(1.27 \frac{\Delta m_{32}^2 L_{\rm 
atm}}{E_{\rm atm}} \right) \nonumber \\ & & \!\! -\; 2 |V_{1 \mu}|^2 
\left( 1 - |V_{1 \mu}|^2 \right)
\end{eqnarray}

\ni and

\begin{eqnarray}
\!\!\!P\left(\nu_\mu \rightarrow \nu_e \right)\!\!\! & \simeq & \!\!\!4 
|V_{1 e}|^2 |V_{1 \mu}|^2 \sin^2 \left( 1.27 \frac{\Delta m_{21}^2 L_{\rm LSND}
}{E_{\rm LSND}} \right) \;, \\ \!\!\!P\left(\nu_\mu \rightarrow \nu_\tau 
\right)\!\!\! & \simeq & \!\!\!4\left( |V_{2 \tau}|^2 |V_{2 \mu}|^2 + 
V_{1 \tau} V_{1 \mu}^* V_{2 \tau}^* V_{2 \mu} \right) \sin^2 \left( 1.27 
\frac{\Delta m_{32}^2 L_{\rm atm}}{E_{\rm atm}} \right) + 2 |V_{1 \tau}|^2 
|V_{1 \mu}|^2 \nonumber \\ & \sim & \!\!\!4 |V_{2 \tau}|^2 |V_{2 \mu}|^2 
\sin^2 \left( 1.27 \frac{\Delta m_{32}^2 L_{\rm atm}}{E_{\rm atm}}\right)\;,
\end{eqnarray}

\ni the last step being valid for the estimate $ |V_{1 \mu}| \sim 0 $ (compare 
Eqs. (27), where $ |V_{1 \mu}| \sim 0.07 $; if the LSND effect does not 
exist, $|V_{1 \mu}|$ ought to be distinctly smaller). From Eqs. (21) and (23) 
with the estimates (25) we can see that atmospheric $\nu_\mu $'s oscillate
dominantly into $\nu_\tau $'s. Similarly, Eqs. (20) and (24) imply that $\nu^{
(a)}_e $'s oscillate dominantly into $\nu^{(s)}_e $'s.

 When comparing Eqs. (20), (21) and (22) with experimental estimates, we obtain
for solar $\nu_e $'s (taking the global vacuum solution) [4]

\begin{eqnarray} 
|V_{1 e}|^4 \leftrightarrow \sin^2 2\theta_{\rm sol} \sim 1\;,\;4 m_1^2 
\lambda^{(M)}\! & \!\!\!\leftrightarrow\!\!\! & \Delta m_{\rm sol}^2 \sim 
10^{-10}\;\;{\rm eV}^2\;,\nonumber \\
\frac{1}{2}\left[|V_{2 e}|^4 + |V_{3 e}|^4 + 4 |V_{1 e}|^2 \left( 1 - \right.
\right.\!\! & \!\!\!\!|V_{1 e}|^2\!\!\!\! & \!\!\!\left. \left. \right) + 4 
|V_{2 e}|^2 |V_{3 e}|^2\right] \nonumber \\
\equiv \frac{1}{2}\left[ \left( 1 + 3|V_{1 e}|^2\right)\left(1 - |V_{1 e}|^2
\right)\right.\!\!\! & \!\!\!\!\!+\!\!\!\! & \!\!\!\left.2 |V_{2 e}|^2 |V_{3 e}
|^2 \right]\!\! \sim 0\,, 
\end{eqnarray}

\ni for atmospheric $\nu_\mu $'s [5]

\begin{eqnarray} 
4|V_{2 \mu}|^2|V_{3 \mu}|^2 \leftrightarrow \sin^2 2\theta_{\rm atm} \sim 1 
& , & \Delta m_{32}^2 \leftrightarrow \Delta m_{\rm atm}^2 \sim 3\times 10^{-3}
\;{\rm eV}^2\;,\nonumber \\ |V_{1 \mu}|^2 \left( 1 \right.\!\! & \!\!\!\! - 
\!\!\!\! & \!\!\left.|V_{1 \mu}|^2 \right) \sim 0
\end{eqnarray}

\ni and for LSND $\nu_\mu $'s [6]


\begin{equation} 
4|V_{1 e}|^2|V_{1 \mu}|^2 \leftrightarrow \sin^2 2\theta_{\rm LSND} \sim 0.02 
\;,\; \Delta m_{21}^2 \leftrightarrow \Delta m_{\rm LSND}^2 \sim 0.5 \;{\rm eV
}^2\;.
\end{equation}

\ni Hence,


\begin{equation} 
|V_{1 e}|^2 \sim 1 \;,\; |V_{1 \mu}|^2 \sim 0.005 \simeq 0 \;,\;4|V_{2 e}|^2 
|V_{3 e}|^2 \sim 0 \;,\; 4|V_{2 \mu}|^2 |V_{3 \mu}|^2 \sim 1 
\end{equation}

\ni and


\begin{equation} 
4 m_{1}^2 \lambda^{(M)} \sim 10^{-10} \;{\rm eV}^2\;,\; m^2_3 - m^2_2 \sim
3\times 10^{-3} \;{\rm eV}^2\;,\; m^2_1 \ll m^2_2 \simeq m^2_3 \sim 0.5\;
{\rm eV}^2\;.
\end{equation}

\ni Thus, Eq. (19) requires


\begin{equation} 
\lambda^{(M)\,2} \ll \left(\frac{3}{2}\right)^2\times 10^{-6} \;.
\end{equation}

\ni On the other hand, $\lambda^{(M)} \sim (1/4 m^2_1)\times 10^{-10}\;{\rm eV
}^2 \gg \frac{1}{2}\times 10^{-10}$.

 Note that for the Chooz experiment [7] on possible deficit of reactor 
$\overline{\nu}_e$'s our oscillation formula (16) (with $\nu_e $ replaced 
by $\overline{\nu}_e$) and numerical conjecture (18) + (19) lead to

\begin{equation} 
P(\overline{\nu}_e \rightarrow \overline{\nu}_e) \simeq 1 - 4 |V_{2 e}|^2 |V_{
3 e}|^2 \sin^2 \left( 1.27 \frac{\Delta m_{32}^2 L_{\rm Chooz}}{E_{\rm Chooz}}
\right) - 2|V_{1 e}|^2\left(1 - |V_{1 e}|^2 \right)\;,
\end{equation}

\ni since $ L_{\rm Chooz}/E_{\rm Chooz} \simeq L_{\rm atm}/E_{\rm atm}$ 
roughly. Thus, its negative result $P(\overline{\nu}_e \rightarrow \overline{
\nu}_e) \sim 1 $ implies $|V_{2 e}|^2|V_{3 e}|^2\sim 0 $ and $|V_{1 e}|^2 \sim 
1 $, consistently with the estimates (27) following from solar experiments.

 Concluding, with the conditions (27), (28) and (29) satisified, oscillations 
of three pseudo--Dirac neutrinos can nicely describe all three neutrino 
experimental effects, without introducing any extra sterile neutrinos. If the 
LSND effect does not exist, the value $|V_{1 \mu}|^2\sim 0 $ ought to be 
distinctly smaller than 0.005.

 The experimental estimates (27) are compatible with the following form [8] of 
neutrino family mixing matrix:


\begin{equation}
V^\dagger = 
\left(\begin{array}{ccc} c  & s  &  0 \\  - s/\sqrt{2}  & c/\sqrt{2}  & - 
1/\sqrt{2} \\ - s/\sqrt{2}  & c/\sqrt{2}  & ~~1/\sqrt{2} \end{array} \right)
\end{equation}

\ni (with $ c = \cos \theta$, $ s = \sin \theta $ and all phases neglected),
corresponding to the maximal mixing of $\nu_\mu $ and $\nu_\tau $ within $
\nu_i^{I,\,II} = \sum_\alpha V_{i \alpha} \nu_\alpha^{I,\,II}$, where $\nu^{I,
\,II}_\alpha \simeq ( \nu_\alpha^{(a)} \mp \nu_\alpha^{(s)})/\sqrt{2}$ ($\alpha
=  e\,,\,\mu\,,\,\tau\;,\;\;i = 1,2,3$). In fact, all experimental conditions 
(27) are then satisfied if $ c^2 \sim 0.99 \simeq 1 $ and $ s^2 \sim 0.01 $ or


\begin{equation} 
c \sim \sqrt{0.99} \simeq 1 \;\;,\;\; s \sim 0.1 \;.
\end{equation}

\ni In this case,


\begin{equation}
V^\dagger \sim \left(\begin{array}{ccc} \sqrt{0.99} & 0.1 & 0 \\ - 0.1/\sqrt{2}
& \sqrt{0.99}/\sqrt{2} & - 1/\sqrt{2} \\ - 0.1/\sqrt{2} & \sqrt{0.99}/\sqrt{2} 
& ~~1/\sqrt{2} \end{array} \right)\;\;.
\end{equation}

\ni If here $ U^{(e)}_{\alpha \beta} \simeq \delta_{\alpha \beta}$, then
$ U^{(\nu)}_{\alpha i} \simeq (V^\dagger)_{\alpha i} = V^*_{i \alpha}$.

 Such a neutrino family diagonalizing matrix $ U^{(\nu)} \simeq V^\dagger $ is 
related to the neutrino family mass matrix $ M^{(\nu)}$ expressed through its 
eigenvalues $ m_i $ ($ i = 1,2,3 $) by the formula [see Eqs. (5)]:


\begin{eqnarray}
\lefteqn{M^{(\nu)} = \left(\sum_i U^{(\nu)}_{\alpha i} U^{(\nu)\,*}_{\beta i}
m_i \right)} \nonumber \\ & & \!\!\simeq \left(\begin{array}{ccc} m_1 c^2 + 
m_2 s^2 & (m_2 - m_1) c s /\sqrt{2} & (m_2 - m_1) c s /\sqrt{2} \\ (m_2 - m_1)c
s/\sqrt{2} & (m_1 s^2 + m_2 c^2 + m_3)/{2}  & (m_1 s^2 + m_2 c^2 - m_3)/{2} \\
(m_2 - m_1) c s /\sqrt{2} & (m_1 s^2 + m_2 c^2 - m_3)/{2} & (m_1 s^2 + m_2 c^2 
+ m_3)/{2} \end{array} \right)\;.
\end{eqnarray}

\ni In the case of mass hierarchy (15) realized as $ m_1 \ll m_2 \simeq m_3 $ 
with $ m_1 \sim 0 $, $(m_2 + m_3)/2 \sim \sqrt{0.5}$ eV and  $ m_3 - m_2 
\sim \sqrt{0.5} \times 10^{-3}$ \rm eV, Eqs. (34) and (32) give


\begin{equation}
M^{(\nu)} /{\rm eV} \sim \sqrt{0.5} 
\left(\begin{array}{ccc} 0.01 & \sqrt{0.005} & \sqrt{0.005} \\ \sqrt{0.005} &
~~0.995 & - 0.0065 \\ \sqrt{0.005} & - 0.0065 & ~~0.995 \end{array} \right)\;,
\end{equation}

\ni since $  c s /\sqrt{2} \sim \sqrt{0.005}\;,\;(1 - c^2)/2 \sim 0.005 $ and 
$(1 + c^2)/2 \sim 0.995 $. If the LSND effect does not appear, the value $|V_{
1 \mu}|^2= s^2/2 \sim 0 $ should be distinctly smaller than 0.005 (or $ s $ 
distinctly smaller than 0.1). In this case,


\begin{equation}
M^{(\nu)} /{\rm eV} \sim \sqrt{0.5} \left(\begin{array}{ccc} 0 & 0 & 0 \\
0 & 1 & - 0.0015 \\  0 & - 0.0015 & 1 \end{array} \right)\;,
\end{equation}

\ni leading to $m^2_1 \sim 0$ and $m^2_{2,3} \sim 0.5(1 \mp 0.003)\;{\rm eV}^2 
\simeq 0.5\;{\rm  eV}^2 $ [of course, they are such by construction, both for
Eqs. (35) and for (36)].

 However, if there is really no LSND effect (and atmospheric $\nu_\mu $'s still
oscillate dominantly into $\nu_\tau $'s), then, perhaps, not the three pseudo%
--Dirac neutrinos and mass hierarchy (15) are the most natural option, but 
rather the three see--saw neutrinos and popular mass hierarchy


\begin{equation}
 m^2_{\nu_1} \simeq  m^2_{\nu_2} \ll  m^2_{\nu_3}\;,
\end{equation}

\ni where now $ m^2_{\nu_2} - m^2_{\nu_1} \leftrightarrow \Delta m^2_{\rm sol}$
and $ m^2_{\nu_3} - m^2_{\nu_2} \leftrightarrow \Delta m^2_{\rm atm}$ (see {\it
e.g.}, Ref. [8]).

 Concluding our comments on the phenomenological texture expressed by Eqs. (33)
and (35), we would like to point out that it differs from the texture model 
described in two last Refs. [3]. There, the LSND scale $\Delta m^2_{\rm LSND}$ 
(if it exists) and the atmo\-spheric scale~~$\Delta m^2_{\rm atm}$ are 
interpreted as $ m^2_3 - m^2_2 $ and a number $\,\stackrel{>}{\sim} 4 m^2_2 
\lambda^{(M)}\,$ (but $\ll m^2_3 - m^2_2 $), respect\-ively. In contrast, 
their present interpretation is $ m^2_2 - m^2_1 $ and $ m^2_3 - m^2_2 $, 
respect\-ively. In both textures the hierarchy (15) holds.

\vfill\eject 

~~~~
\vspace{0.6cm}

\centerline{\bf References}

\vspace{1.0cm}

{\everypar={\hangindent=0.5truecm}
\parindent=0pt\frenchspacing

{\everypar={\hangindent=0.5truecm}
\parindent=0pt\frenchspacing

~1.~M.~Gell--Mann, P.~Ramond and R.~Slansky, in {\it Supergravity }, ed. 
P.~van Nieuwenhuizen and D.Z.~Freedman, North--Holland, Amsterdam, 1979;
T.~Yanagida, in {\it Proc. of the Workshop on the Unified Theory of the 
Baryon Number in the Universe}, ed. O.~Sawada and A.~Sugamoto, KEK report 
No. 79--18, Tsukuba, Japan, 1979; see also R.~Mohapatra and G.~Senjanovic,
{\it Phys. Rev. Lett.} {\bf 44}, 912 (1980).

\vspace{0.15cm}

~2.~For a recent discussion see C. Giunti, hep--ph/9908513; and references 
therein.

\vspace{0.15cm}

~3.~D.W.~Sciama, astro--ph/9811172; and references therein;  A. Geiser, 
CERN--EP/98--56, hep--ph/9901433;  W. Kr\'{o}likowski, hep--ph/9903209, {\it 
Acta Phys. Pol.} {\bf B} (Appendix), to appear; {\it Nuovo Cimento} {\bf A}, 
to appear. 

\vspace{0.15cm}

~4.~See {\it e.g.}, J.N. Bahcall, P.I. Krastov and A.Y. Smirnov, hep--ph/%
9807216v2.

\vspace{0.15cm}

~5.~Y. Fukuda {\it et al.} (Super--Kamiokande Collaboration), {\it Phys. Rev. 
Lett.} {\bf 81}, 1562 (1998); and references therein. 

\vspace{0.15cm}

~6.~C.~Athanassopoulos {\it et al.} (LSND Collaboration), {\it Phys. Rev.}
{\bf C 54}, 2685 (1996); {\it Phys. Rev. Lett.} {\bf 77}, 3082 (1996); nucl--%
ex/9709006.

\vspace{0.15cm}

~7.~M.~Appolonio {\it et al.} (Chooz Collaboration), {\it Phys. Lett.}
{\bf B 420}, 397 (1998).

\vspace{0.15cm}

~8.~G.~Altarelli and F.Feruglio, CERN--TH/99--129 + DFPD--99/TH/21, hep--ph/%
9905536; and references therein.

\vfill\eject

\end{document}